\documentstyle[pre,aps,preprint,eqsecnum,epsfig]{revtex} 

\tightenlines

\newcommand{\app}{\rightarrow}

\newcommand{\bigpb}{P_B}
\newcommand{\bigt}{T_{\!\dt}}

\newcommand{\dimscite}{Pratt-1986,Theodorou-1993,Chandler-1998a,Woolf-1998e,Chandler-1998b,Chandler-1998c,Chandler-1998d,Chandler-1998e,Mazonka-1998,Zuckerman-1999,Eastman-2000}
\newcommand{\dee}{\partial}
\newcommand{\dt}{\Delta t}
\newcommand{\dxr}{\Delta x_R}

\newcommand{\omj}{{\mathrm{OMJ}}}
\newcommand{\om}{{\mathrm{OM}}}
\newcommand{\qtilde}{\tilde{Q}}
\newcommand{\sigsim}{\sigma_k}
\newcommand{\som}{{\cal L}_{\om}}
\newcommand{\somj}{{\cal L}_{\omj}}
\newcommand{\totd}{{\mathrm{d}}}

\newcommand{\xdot}{\dot{x}}

\newcommand{\xth}{x_h}

\begin{document}
\title{Efficient Dynamic Importance Sampling of Rare Events in One Dimension}
\author{Daniel M. Zuckerman$^\ast$ and Thomas B. Woolf$^{\ast \dagger}$}
\address{$^\ast$Department of Physiology and  $^\dagger$Department of Biophysics,\\
Johns Hopkins University School of Medicine, Baltimore, MD 21205\\
\texttt{dmz@groucho.med.jhmi.edu, woolf@groucho.med.jhmi.edu}
}
\date{\today}
\maketitle

\begin{abstract}
Exploiting stochastic path integral theory, we obtain \emph{by simulation} substantial gains in efficiency for the computation of reaction rates in one-dimensional, bistable, overdamped stochastic systems.
Using a well-defined measure of efficiency, we compare implementations of ``Dynamic Importance Sampling'' (DIMS) methods to unbiased simulation.
The best DIMS algorithms are shown to increase efficiency by factors of approximately 20 for a $5 k_B T$ barrier height and 300 for $9 k_B T$, compared to unbiased simulation.
The gains result from close emulation of natural (unbiased), instanton-like crossing events with artificially decreased waiting times between events that are corrected for in rate calculations.
The artificial crossing events are generated using the closed-form solution to the most probable crossing event described by the Onsager-Machlup action.
While the best biasing methods require the second derivative of the potential (resulting from the ``Jacobian'' term in the action, which is discussed at length), algorithms employing solely the first derivative do nearly as well.
We discuss the importance of one-dimensional models to larger systems, and suggest extensions to higher-dimensional systems.
\end{abstract}
\pagebreak

\begin{figure}
\begin{center}
\epsfig{file=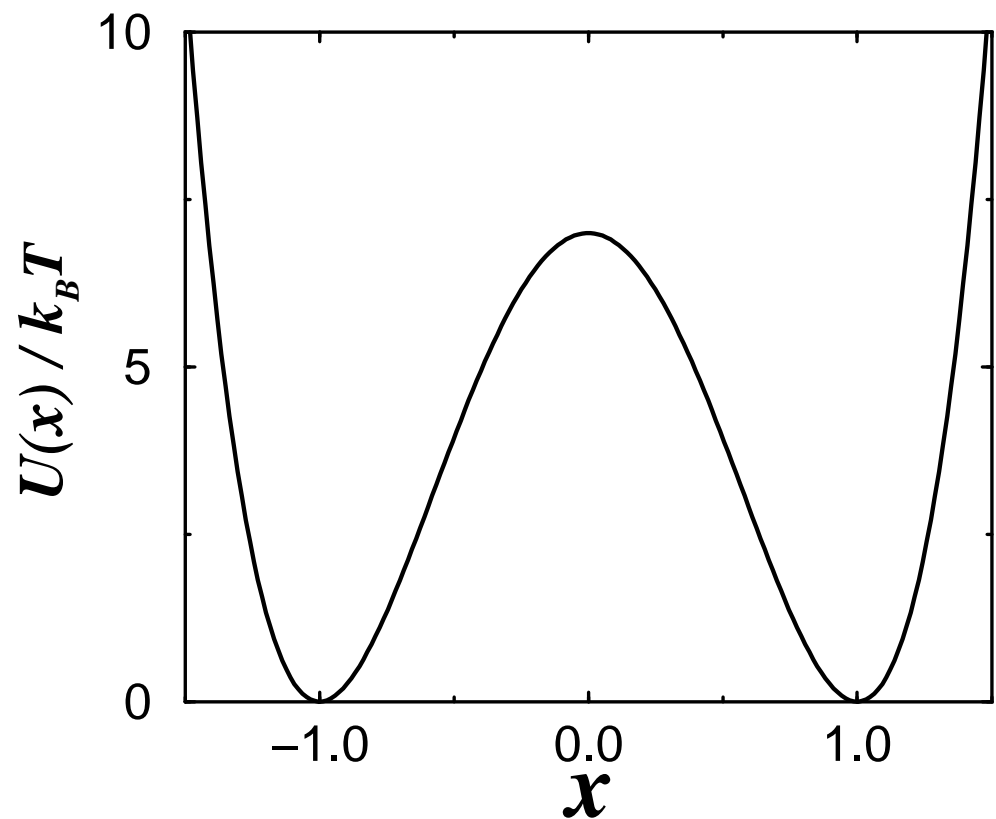, height=4in}
\end{center}
\vspace*{1cm}
\caption{\label{fig:potl}
The symmetric bistable potential studied in the text, Eq.\ (\ref{potl}), shown for a barrier height of $E_b = 7 k_B T$ and length parameter $l=1$, the unit of the $x$ axis.
}
\end{figure}
\pagebreak

\begin{figure}
\begin{center}
\epsfig{file=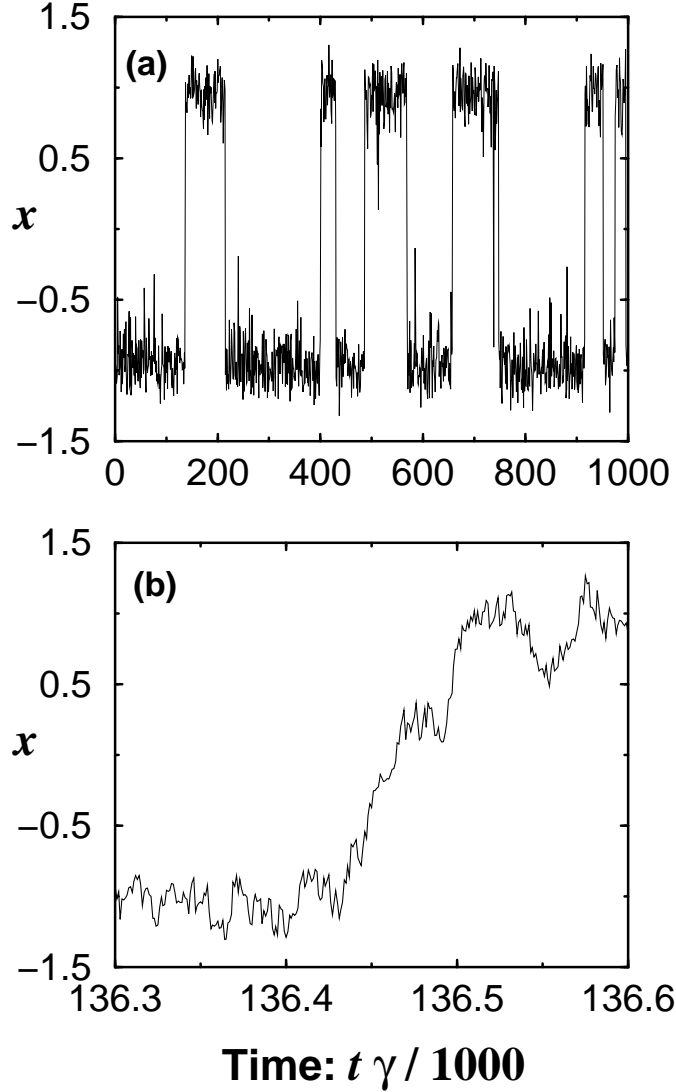, height=5.5in}
\end{center}
\vspace*{1cm}
\caption{\label{fig:time}
Time scales and crossing events.
An unbiased trajectory exhibits two vastly different time scales: for a substantial barrier height, the waiting time between events greatly exceeds the time for a single crossing event.
(a) The top plot shows an unbiased trajectory of an overdamped particle in the double well of Eq.\ (\ref{potl}), with a barrier height of $E_b = 7 k_B T$.
(b) The bottom plot isolates a single crossing event from the trajectory in (a) and highlights the rapidity (steepness) of the crossing, including of the ascent.
Note that the unit of $x$ in both plots is the lengthscale $l$; see Eq.\ (\ref{potl}).
}
\end{figure}
\pagebreak

\section{Introduction}
The rapid computation of the transition rate \emph{by numerical simulation} for an overdamped stochastic particle confined to a one-dimensional double-well potential (Fig.\ \ref{fig:potl}) is a deceptively simple problem, and of crucial importance to making progress in multi-dimensional systems.
While straightforward unbiased simulation ultimately yields the rate to any desired precision (e.g., \cite{Pastor-1994b}), it requires the simulator to endure many times the waiting period \emph{between} the rare transition events (Fig.\ \ref{fig:time}a).
Such unbiased simulation, moreover, is utterly impracticable for larger systems
--- particularly biomolecules possessing thousands of atoms, for which simulating a waiting time of 1 millisecond could require tens or hundreds of centuries of computer time.
The well-known analytic methods for rate computations in simple systems \cite{Hanggi-1990} are also insufficient for high-dimensional, rough energy landscapes because
many barriers and metastable states of unknown geometry intercede along multiple unknown pathways between the two states of interest (see, e.g., \cite{Chandler-1998f,Chandler-1998g}).

The inadequacy of both analytic methods and straightforward simulation for large biochemical systems points directly to the need for importance sampling and related methods \cite{\dimscite,Straub-1997,Ruiz-1997,Shalloway-1998}.
Importance sampling techniques focus computational effort on transition events, typically generating an ensemble of transition trajectories from which to estimate rates and/or paths.
Yet despite the early successes and formal appeal of these methods, a quantitatively successful computational tool for large protein systems with $10^4$ atoms has not been achieved.
We believe
the apparently trivial bistable one-dimensional system must be completely understood \emph{in a simulation context} before substantial progress can be expected for importance sampling in larger systems.

The present paper is explicitly computational, or ``simulational'':
the sole objective is to develop efficient simulation methods for one-dimensional systems which are directly or indirectly applicable to large systems.
We concern ourselves only with the ``dynamic importance sampling'' (DIMS) methods developed by Woolf \cite{Woolf-1998e} and Zuckerman and Woolf \cite{Zuckerman-1999}, which generate ensembles of fully independent transition trajectories.
We hope and believe our results will be of practical use in other multi-dimensional methods, but such applications are beyond the scope of this work.
Rather, the ``bottom-line'' questions we attempt to answer are:
(i) \emph{
Within the DIMS framework, what are the most efficient methods for rate computation?}
(ii) \emph{
Using a well-defined measure, how efficient are these methods compared to unbiased simulation?}
We will also discuss the maximum efficiency possible using DIMS and related methods in one-dimensional (1D) problems, as well as attempting to extrapolate to larger systems.

\begin{figure}
\begin{center}
\epsfig{file=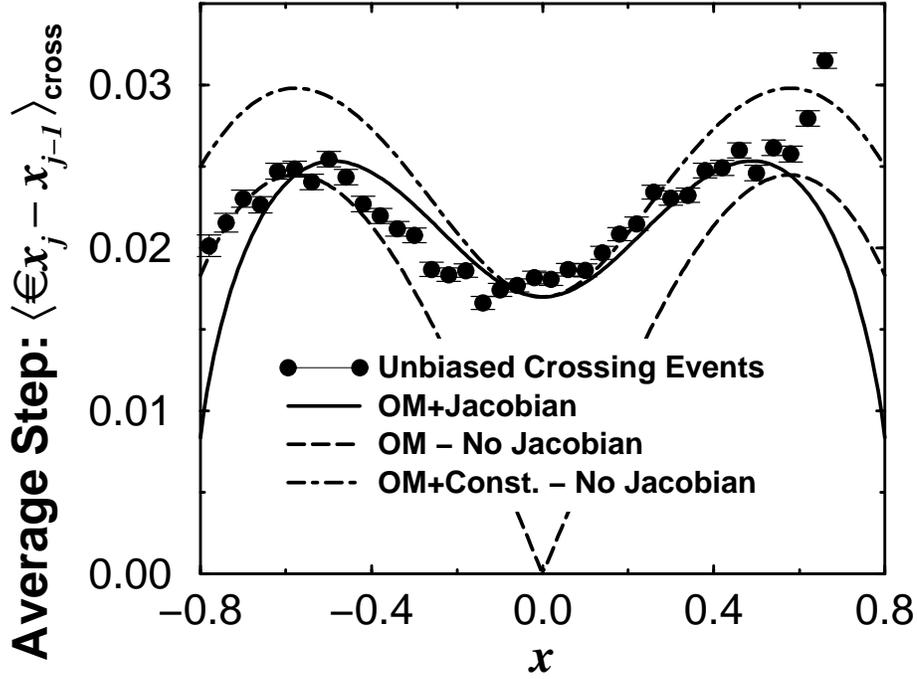, height=4in}
\end{center}
\caption{\label{fig:cross_eb7}
Average simulation step sizes ($x_j - x_{j-1}$) from many unbiased crossing events generated by Eq.\ (\ref{sim}).
A typical crossing event is shown in Fig.\ \ref{fig:time}(b).
Solid circles are the data averaged from \emph{unbiased crossing events only} for the potential of Eq.\ (\ref{potl}), with a barrier height $E_b = 7 k_B T$. 
The solid line is the most probable path Eq.\ (\ref{xcomj}) based on the Onsager-Machlup action with the stochastic correction (``Jacobian'') term; 
the dashed lines depict the uncorrected Onsager-Machlup extremal path, Eq.\ (\ref{xcom}) with $C=0$;
and, the dot-dash lines show the uncorrected Onsager-Machlup path with a non-vanishing constant --- i.e., Eq.\ (\ref{xcom}) with $C=- 2 k_B T \, U''(0)/(m \gamma)^2$.
The statistical error in the simulation data may be gauged by the noise in the data.
The data for $|x| > 0.6$ are affected ``artifactually'' by the procedure of extracting crossing events from long trajectories; see the text, however.
The data of this figure only were generated using the parameters $\dt = 10^{-4} \gamma^{-1}$, $\gamma^{-1}\equiv1$, $m=10.98$, and $k_B T =  249.462$, largely following Refs.\ [8,14]; 
all lengths are given in units of the lengthscale $l$ of Eq.\ (\ref{potl}).
}
\end{figure}

Achieving efficiency for a 1D problem using dynamic importance sampling, as we will show, requires the appreciation of a variety of theoretical results --- particularly relating to stochastic path integrals.
The review by Mortensen \cite{Mortensen-1969} and
Gardiner's book \cite{Gardiner-1985} give introductions to stochastic integrals --- of which Ito's and Strantonovich's are the most basic types.
Historically, Onsager and Machlup first formulated a stochastic path integral to describe overdamped Brownian dynamics \cite{Onsager-1953a}.
Since then, uniqueness and other formal aspects of the Onsager-Machlup formulation have been discussed by a variety of workers (e.g., \cite{Stratonovich-1962,Haken-1976,Graham-1976,Bach-1977,Wissel-1979,Caroli-1981,Dykman-1994}).
Indeed, formal properties of stochastic functional integrals also play an important role in the present investigation.
These integrals have been used, for example, to address the question of the most likely crossing event (see Figs.\ \ref{fig:time}b, \ref{fig:cross_eb7}) \cite{Graham-1975,Dykman-1979,Bray-1989,Dykman-1990,McKane-1990a,McKane-1990b,McKane-1990c,McKane-1993,Elber-1997,Astumian-1999,Elber-2000}.  
Related functional integral approaches directly address the Smoluchowski (overdamped Fokker-Planck) equation for the evolution of the entire probability distribution \cite{Stratonovich-1962,Graham-1973,Graham-1975,Weiss-1982,Baibuz-1984}.  

Path integral formulations lead naturally to the notion of an average path, which is critical to the present discussion --- even in one dimension.
While the geometric pathway for barrier crossing is trivial in a one-dimensional potential like that of Fig.\ \ref{fig:potl}, the ``speed'' (average displacement) at every position constitutes another dimension of the average path --- and a critical one to the present discussion.
Such 1D average paths (average speeds and distributions) were considered by Dykman, McClintock and coworkers \cite{Dykman-1992} and by Luchinsky and McClintock \cite{McClintock-1997}, both theoretically and experimentally.
Zuckerman and Woolf \cite{Zuckerman-1999} considered average paths in higher dimensions, computationally, while the problem of finding optimal reaction paths in multi-dimensional systems has a long history (e.g., \cite{Muller-1979,Elber-1990,Elber-1991,Karplus-1992,Theodorou-1993,Elber-1997}.

The paper is organized as follows.
We first review the basic formalism for overdamped Langevin dynamics and importance sampling in Section \ref{sec:back}.
Section \ref{sec:mpp} introduces pertinent path-integral results based on the Onsager-Machlup action, and reviews the so-called ``Jacobian'' term in some detail, as well as presenting new numerical data.
Section \ref{sec:bias} discusses a new method for producing an efficient, biased ensemble of trajectories, while Section \ref{sec:results} gives the actual results from the new biasing technique.
Section \ref{sec:multi} addresses the relevance of our results to multidimensional problems.
Finally, conclusions and a summary of the results are given in Section \ref{sec:sum}.


\pagebreak
\section{Overdamped Langevin Dynamics and Importance Sampling}
\label{sec:back}
This section briefly reviews the fundamental stochastic dynamics equations and the dynamic importance-sampling formalism for rate computations.
We consider solely stochastic dynamics governed by the overdamped Langevin equation in the presence of Gaussian white noise.
In the notation of \cite{Zuckerman-1999}, such ``Brownian dynamics'' are described by
\begin{equation}
\label{brownian_eom}
\totd x /\totd t = f / m \gamma + R(t),
\end{equation}
where $x$ is the configurational coordinate, $t$ is time, $f(x) = - \nabla U(x)$ is the force, $m$ is the particle's ``mass'', $\gamma$ is the friction constant, and the noise $R$ is assumed Gaussian with zero mean and variance given according to
\begin{equation}
\langle \, R(t) \,R(t^\prime) \, \rangle = ( 2 k_B T / m \gamma ) \, \delta(t-t^\prime) \, ,
\end{equation}
where $k_B$ is Boltzmann's constant and $T$ the temperature.
The equation (\ref{brownian_eom}) is presumed to be simulated according to the ``Ito-like'' (Euler) discretization \cite{Allen-Tildesley,Kloeden-1992}
\begin{equation}
\label{sim}
x_{j+1} = x_j + (f_j / m \gamma) \dt + \dxr ,
\end{equation}
where the subscripts $j$ indicate quantities evaluated at time $j \dt \equiv t_j$, so that $f_j = f(x_j)$, and $\dxr$ is chosen from a Gaussian distribution of zero mean and variance
\begin{equation}
\label{natural_variance} 
\sigma^2 = 2 \dt k_B T/m \gamma \;.
\end{equation}
The discretization (\ref{sim}) is considered to be Ito-like because the force --- assumed to be constant over the interval $\dt$ --- is evaluated at the beginning of the interval;
a loosely-termed ``Stratonovich-like'' approach would instead consider, formally \cite{Graham-1975,Graham-1976,Gardiner-1985},
\begin{equation}
\label{straton}
f_j \app (f_j + f_{j+1})/2 \; .
\end{equation}

We note that discretizations more sophisticated than the Euler scheme (\ref{sim}) are well known --- i.e., Runge-Kutta schemes of various orders \cite{Rumelin-1982,Batrouni-1985,Branka-1998}.
However, in explicit tests on \emph{rate calculations only} for our simple system (Fig.\ \ref{fig:potl}), the lowest-order Runge-Kutta (Heun) method permitted only a marginally larger time step --- without losing accuracy --- which did not justify the additional computational expense.  
Larger time steps tend to produce \emph{systematic} errors, in addition to the easier-to-diagnose statistical error, regardless of the algorithm \cite{Branka-1998}.
Rate calculations, moreover, may be especially sensitive.
We also note that methods of higher-order than the Heun algorithm are not of interest here, because computing higher-order derivatives is extremely costly for the larger biomolecular systems toward which the present research is ultimately oriented.
Thus, given our focus on rate computations and our interest in large systems, there seems little reason to employ anything more complicated than the Euler procedure (\ref{sim}).

It is useful to consider, in parallel to the dynamical algorithm, the equivalent probabilistic description of the dynamics (\ref{sim}).
In particular, the explicit single-step transition probability density described above is given by
\begin{equation}
\label{trans_prob_density}
\bigt(x_{j+1}|x_j) = ( 2 \pi \sigma^2 )^{-1/2} 
	\exp{\left[ - \left[ \, (x_{j+1} - x_j) -  (f_j / m \gamma) \dt \, \right]^2 / 2 \sigma^2 \right] } \; .
\end{equation}
Thus the probability density for a whole trajectory $\zeta = (\,x_0, \,x_1, \,x_2, \ldots, \,x_n)$ is given by the product of the single-step densities:
\begin{equation}
\label{qtilde}
\qtilde(\zeta) = \prod_{j=0}^{n-1} \bigt(x_{j+1} | x_j ).
\end{equation}

Importance sampling is effected \cite{Woolf-1998e,Zuckerman-1999} by performing \emph{biased} simulations which follow a prescription different from (\ref{sim}).
Naturally, trajectories generated from a biased simulation are distributed not according to $\qtilde$, but according to some other function, $D(\zeta)$.
This alternative distribution can be used to compute probabilities and, hence, transition rates.
One first requires the conditional probability $\bigpb(t|0;x_0)$ to be in the final state (``B'') at time $t$ having started at $x_0$ at $t=0$, which can be written in two equivalent ways for non-vanishing $D$:
\begin{equation}
\label{pbint}
\bigpb(t_n|0;x_0) 
\simeq \int \totd \zeta \, \qtilde(\zeta) \, h_B(x_n)\,
  =  \int D(\zeta) \totd \zeta \, \frac{ \qtilde(\zeta)}{D(\zeta)} \, h_B(x_n),
\end{equation}
where $\totd \zeta = \prod_{j=1}^n \totd x_j$, and $h_B(x)$ is an indicator function which is unity for $x$ in state B and zero otherwise.
The approximation here is simply the neglect of the discretization error.
In a biased simulation, the integral is estimated by 
\begin{equation}
\label{pbsum}
\bigpb(t_n|0;x_0) 
  \simeq \frac{1}{M} \sum_{\zeta \in S_D} \frac{ \qtilde(\zeta) }{D(\zeta)} \, h_B(x_n)
\end{equation}
where $S_D$ is a sampling ensemble of $M$ trajectories chosen according to $D$, and where normalization, $\int \totd \zeta D(\zeta) = 1$, is required.
Note that the function $D$ must be known for the simulation to be performed, just as (\ref{trans_prob_density}) and (\ref{qtilde}) are the distributions from which unbiased steps and trajectories are sampled.
Ultimately, rates are estimated by computing the slope of the linear regime in a probability vs. time plot, as discussed in Sec.\ \ref{sec:bias}.

\newpage

\pagebreak
\section{Path Integrals and Most Probable Crossing Events}
\label{sec:mpp}

Achieving efficiency for a 1D problem using dynamic importance sampling requires a variety of theoretical results for stochastic path integrals and their ``instantons.''
These approaches have been used, for example, to address the question of the most probable crossing event (see Figs.\ \ref{fig:time}b, \ref{fig:cross_eb7}), which can be derived in a straightforward way from the continuum Onsager-Machlup action used in the path integral formulation \cite{Graham-1975,Bray-1989,McKane-1990a,McKane-1990b,McKane-1990c,McKane-1993,Elber-1997,Astumian-1999,Elber-2000}.
%
In this section, we discuss the Onsager-Machlup action and its optimization at somewhat greater length than is required for the computational aims of this paper. 
This is because, beyond the theoretical interest in the action and especially in the so-called ``Jacobian'' term, some older literature deserves a fresh look and detailed discussion.

The path-integral story begins with Onsager and Machlup \cite{Onsager-1953a} and their well-known action for weighting continuum representations of overdamped Brownian trajectories.
In that formulation, a \emph{continuous} trajectory --- presumably intended to represent a smoothed stochastic trajectory --- is weighted according to 
 
\begin{eqnarray}
\label{prob-action}
\mathrm{probability} & \propto & \exp{\left[ -\frac{S}{ (k_B T/m\gamma) } \right]} \mbox{ \hspace{3em} where} \\
\label{action}
S & = & \mbox{$\frac{1}{4}$} \int_{t_i}^{t_f} \totd t \; {\cal L}(t), \mbox{ \hspace{3em} and} \\
\label{som}
{\cal L}_{\om}(t; x, \xdot) & = & \left[ \xdot - \frac{ f(x) }{ m \gamma } \right]^2,
\end{eqnarray}
and where $\xdot = \totd x / \totd t$ and $f(x) = -\totd U / \totd x$.
In their original paper \cite{Onsager-1953a}, Onsager and Machlup applied the Euler-Lagrange equation for functional minimization,
\begin{equation}
\label{euler}
\frac{ \dee {\cal L} }{ \dee x } 
  - \frac{ \totd }{ \totd t } \frac{\dee {\cal L}}{ \dee \xdot } = 0 \;,
\end{equation}
to the case of a single harmonic well.

Subsequent contributions amended the original Onsager-Machlup formulation of the action, (\ref{action}) and (\ref{som}).
First, Stratonovich realized in 1962 \cite{Stratonovich-1962} that if one starts from the product of discrete-step probability densities (\ref{qtilde}) and determines the continuum limit in a rigorous fashion, an additional term arises in the integral representation of the action (\ref{action}), so that instead of (\ref{som}), the effective Lagrangian becomes
\begin{equation}
\label{somj}
\somj = \som + \frac{2 k_B T}{(m\gamma)^2} \, U''(x)
 \; .
\end{equation}
This result was confirmed by Bach \emph{et al.} \cite{Bach-1977}.
In other words, the original description by Onsager and Machlup was incomplete for simulations performed according to the Ito-like algorithm (\ref{sim}).
Note that this new term is proportional to the curvature of the potential and is dominant at a barrier top, for any non-zero noise amplitude ($T>0$).

In the mid 1970s Graham \cite{Graham-1973,Graham-1975} re-derived the term as a Jacobian resulting from changing variables from the fluctuation coordinate to the configuration coordinate --- assuming a Stratonovich-like discretization (\ref{straton}).
Thereafter, apparently, the term has been viewed as a Jacobian, although the propriety of that name in a practical \emph{simulation} context is questionable since, to the authors' knowledge, only the Ito discretization (\ref{sim}) is usable precisely, and that results in a constant Jacobian.
More appropriate terminology might be ``stochastic correction term.''

Both Stratonovich and Graham indicated, importantly, that the inclusion of the stochastic-correction/``Jacobian'' term led to a path-integral representation of the probability distribution consistent with the overdamped Fokker-Planck (Smoluchowski) equation.

Graham also derived the most probable crossing event associated with the corrected action using the Euler-Lagrange equation (\ref{euler}), namely \cite{Graham-1975},
\begin{equation}
\label{xcomj}
\left[ \xdot_c^\omj \right]^2
  = \left[ \frac{ f(x) }{m\gamma} \right]^2 
	- \frac{2 k_B T}{(m \gamma)^2} U''(x) + C \; ,
\end{equation}
where $C$ is a constant of integration.
This result is called an ``instanton'' by field theorists (e.g., \cite{Coleman-1979}) because it describes a rapid transition, as illustrated in Fig.\ \ref{fig:time}.
To appreciate this, note that by taking either the positive or negative square root for $\xdot_c^\omj$ in (\ref{xcomj}), the velocity of ascent is seen to be the same as for descent!
Yet this feature was apparently not appreciated until later; see below.
It will, however, prove central to our goal of obtaining efficiency in simulations.
Below, we also address the computational cost associated with computing a second derivative of the potential.

It is worth noting that although the extremal path, formally, results when $C<0$ in (\ref{xcomj}) [from minimizing (\ref{action}) with (\ref{somj})], one cannot necessarily take the constant to be negative or even to vanish.
Observe that the left-hand side of (\ref{xcomj}) must be positive.
Thus, in regions of relatively large positive curvature --- such as near a well bottom --- the second term of the right-hand side could exceed the first in magnitude, requiring $C>0$.
While this observation appears to be immaterial to our investigation of the simple bistable well (see Fig.\ \ref{fig:cross_eb7}), one could imagine the positivity of the constant having physical consequences in a more complicated potential with metastable intermediate states.

Other features of the most probable crossing have also proved of great interest.
Dykman and Krivoglaz made important observations in their study of transitions in nonequilibrium systems using an uncorrected Onsager-Machlup functional integral, in 1979 \cite{Dykman-1979}.
In a series of papers begun in 1989 and aimed primarily at systems with colored noise, Bray, McKane and coworkers \cite{Bray-1989,McKane-1990a,McKane-1990b,McKane-1990c,McKane-1993} also provided some insights pertinent to the case of white noise \emph{without} the correction term.
(While the latter group included the Jacobian term formally, it was omitted from their analysis of the low-temperature limit.)
Both groups derived essentially a special case of Graham's result (\ref{xcomj})
\begin{equation}
\label{xcom}
\left[ \xdot_c^\om \right]^2 = \left[ \frac{ f(x) }{m\gamma} \right]^2 + C,
\end{equation}
where $C$ is again a constant of integration which vanishes for the most probable case.
It also is equivalent to Graham's result at inflection points of the potential.
Bray, McKane and coworkers noted the crucial fact mentioned above: 
for systems with detailed balance (i.e., time-invariant potentials), the ascent described by $\xdot_c^\om$ is equally rapid as, and symmetric with, the descent (see Fig.\ \ref{fig:time}b).
This point is also implicit in the work of Dykman and Krivoglaz.
Furthermore, both groups recognized that paradoxically the extremal trajectory ($C=0$) \emph{never occurs} \cite{Dykman-1979,McKane-1990c} because an infinite amount of time is required to reach (or descend from) a parabolic barrier top!
This may be seen by computing the barrier crossing time,
\begin{equation}
t_b = \int_{t_i}^{t_f} \totd t = \int_{x_i}^{x_f} \frac{ \totd x }{\xdot_c} \; .
\end{equation}

Such apparently formal observations belie the fact that optimal paths have proven useful in a variety of recent studies.
Dykman \emph{et al.} determined weighted \emph{distributions} of paths (including the optimal) leading to large, non-transitioning fluctuations in a time-oscillatory potential \cite{Dykman-1996}.
Olender and Elber \cite{Elber-1997} discussed the Onsager-Machlup extremal barrier crossing, and employed a discrete version in developing an algorithm for steepest-descent path finding between known initial and final states.
Bier \emph{et al.} \cite{Astumian-1999} discussed extremal crossings in the context of rate calculations for fluctuating barriers.
Elber and Shalloway suggested a temperature dependence for the integration constant of (\ref{xcom}) to overcome the divergent-crossing time difficulty,
and successfully distinguished optimal paths for different regimes of an effective temperature parameter \cite{Elber-2000}.

Both in order to confirm the analytic results reviewed above and to gather data that could usefully inform our efforts to gain efficiency in rate computations, we studied numerically generated crossing events.
We performed very long unbiased simulations of a particle moving according to (\ref{sim}) in a one-dimensional bistable well (Fig.\ \ref{fig:potl}) and ``snipped out'' a large number of crossing events like the one shown in Fig.\ \ref{fig:time}b.
Binning the observed step sizes of \emph{only} the crossing events according to the $x$ position, Fig.\ \ref{fig:cross_eb7} shows that the average step sizes 
$\langle x_j - x_{j-1} \rangle_{\mathrm{cross}}$
closely follow the prediction (\ref{xcomj}) which includes the stochastic correction term.
The two theoretical predictions lacking the correction term --- namely (\ref{xcom}) with two values for $C$ --- appear far less adequate by contrast.
We note that the binned distributions appear to be highly Gaussian so that the average and most probable values essentially coincide.

A comment on the data for $|x| \gtrsim 0.6$ in Fig.\ \ref{fig:cross_eb7} is in order.
The substantial turn-up of the data for $x > 0.6$ is certainly an artifact of isolating crossing-events: by definition, the final step or steps are right-moving.
Thus, if one defines crossing events to end when the value $0.7$ is exceeded, rather than $0.8$ as in the figure, the turn-up occurs for correspondingly smaller $x$ values.
Interestingly, while the same argument applies to the left side of the figure, the data are not similarly affected.
Indeed, the asymmetry between the left and right edges of the plot is striking.
While, at present, we can offer no convincing explanation, it is worth noting that the ``OM+Jacobian'' description becomes less than credible as $x$ increases beyond the right inflection point of the potential, $x \simeq 0.58$.
That description, (\ref{xcomj}) with $C=0$, suggests counter-intuitively that a particle which just completed a transition should fall more slowly than the drift velocity (effectively given by ``OM -- No Jacobian'' for $x>0$ in Fig.\ \ref{fig:cross_eb7}).
Yet it is not clear to the present authors precisely why the applicability of the corrected OM theory should be limited to the region between the inflection points;
intuitively, of course, the inflection points roughly mark the boundaries between the stable states and the barrier-top transition rate.

\pagebreak
\begin{figure}
\begin{center}
\epsfig{file=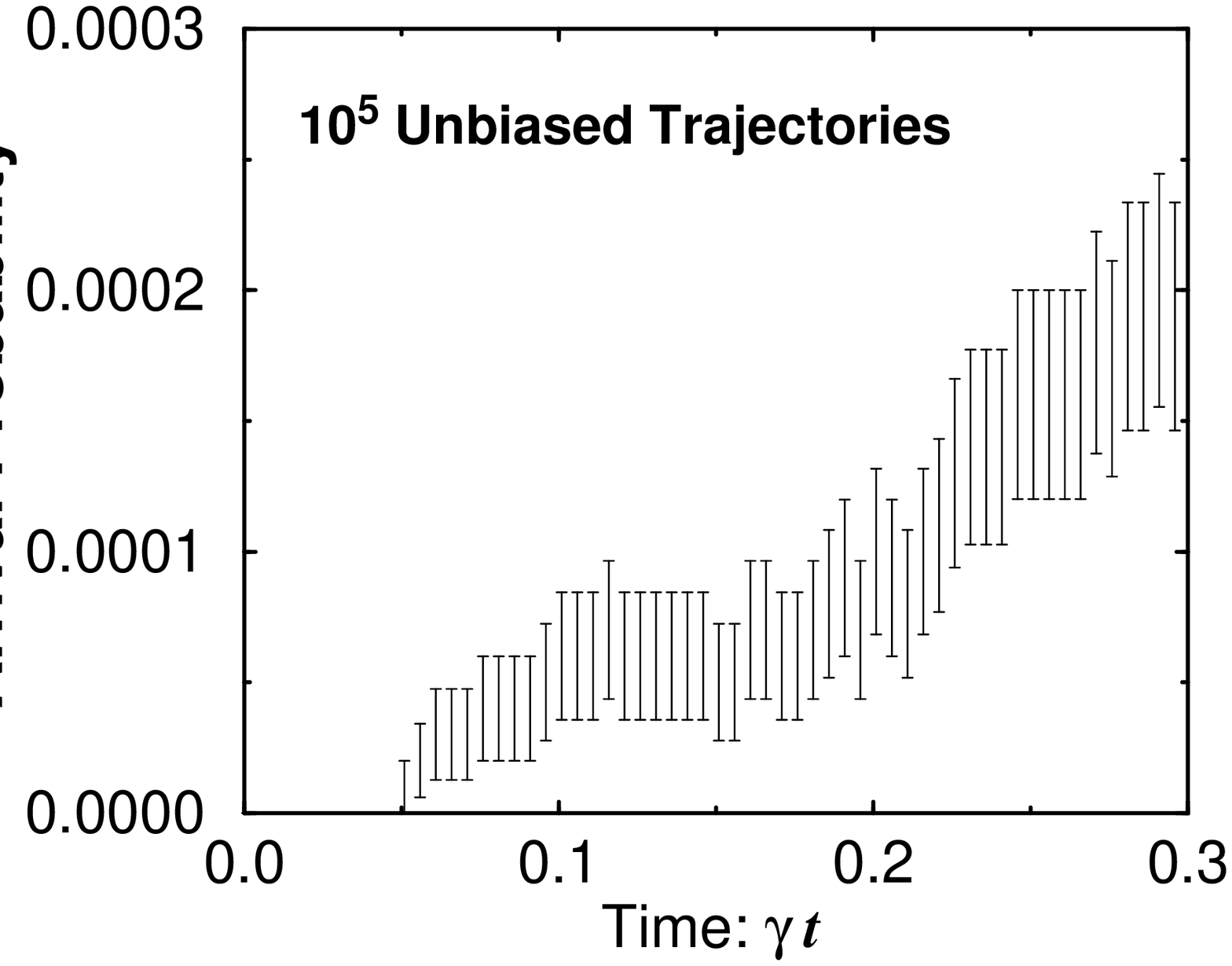, height= 3in}
\epsfig{file=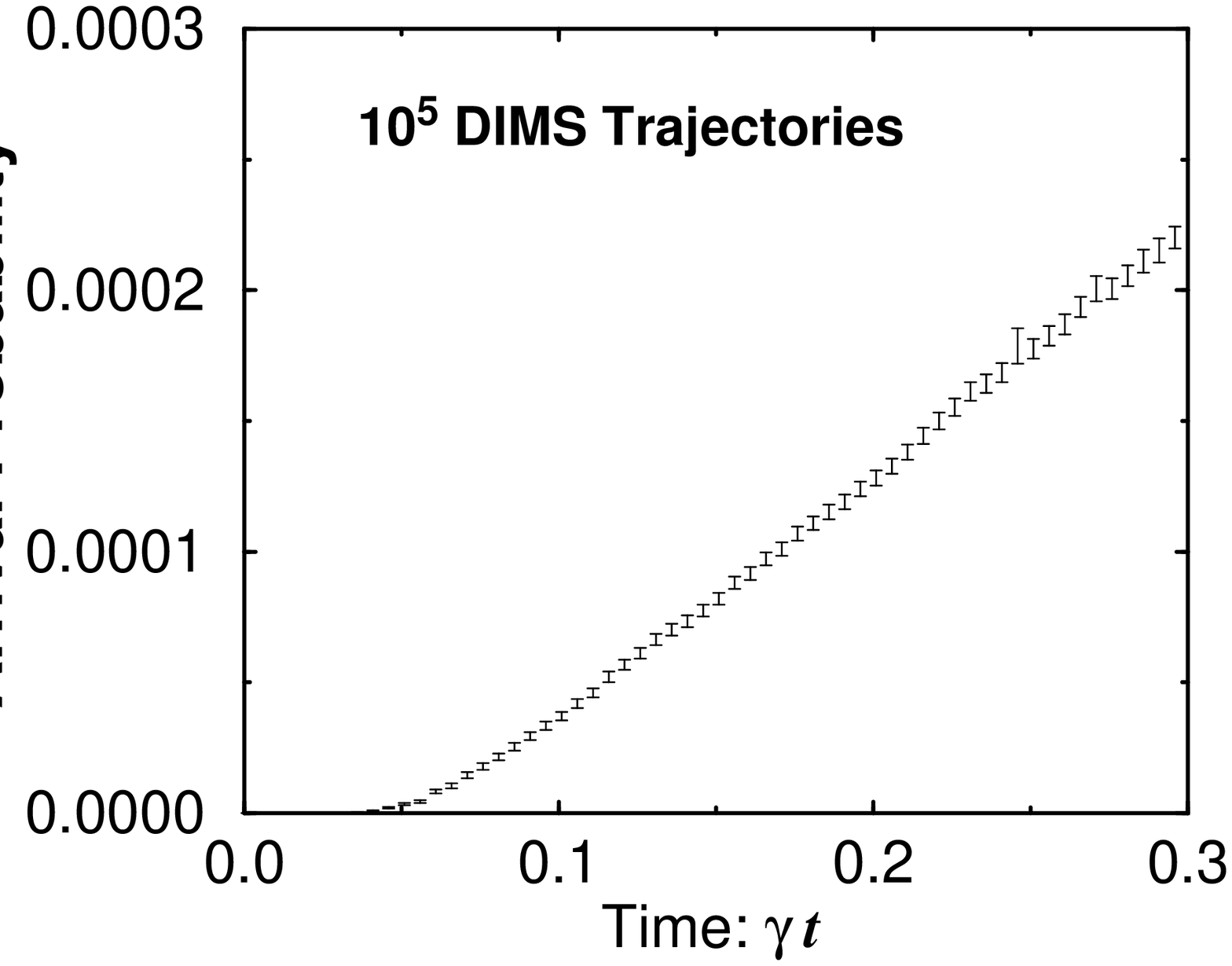, height=3in}
\end{center}
\caption{\label{fig:evo}
Evolution of end-state arrival probability for rate calculations.
The probability to arrive in the right well of the potential Eq.\ (\ref{potl}) with $E_b = 9 k_B T$, depicted in Fig.\ \ref{fig:potl}, is plotted against time, based on trajectories initiated at the bottom of the left well ($x=-1$) at time $t=0$.
Both unbiased (top) and DIMS (bottom) results are shown, with the latter using Eqs.\ (\ref{mpcsim}) and (\ref{xcomj}).
The rate, $k$, is computed as a fit to the slope of the linear regime.
The DIMS computation (bottom) shows a dramatic improvement in efficiency, which is quantified in Fig.\ \ref{fig:effic}.
}
\end{figure}
\pagebreak
\section{A Highly Efficient Bias Method for DIMS Calculations}
\label{sec:bias}
The primary goal of the present work is to improve and quantify the level of efficiency in one-dimensional rate calculations.
To that end we now introduce and evaluate a new biasing method, which is a variation on those discussed in our earlier work \cite{Woolf-1998e,Zuckerman-1999}.
The method combines two essential ingredients: biased crossing events which emulate the most probable path, together with a threshold at which the biasing is triggered.

\subsection{Motivation}
\label{sec:mot}
Computing reaction rates by simulation requires the generation of an ensemble of appropriately weighted trajectories (or a correspondingly long trajectory) exhibiting many barrier-crossing events.
While many sampling methods are capable of generating a suitable ensemble of trajectories in a long period of time, it is no easy task to rapidly generate a truly \emph{important} ensemble --- containing highly probable crossing events --- even for an apparently trivial potential like a symmetric, one-dimensional double well.
The reason is not hard to see.
In an unbiased simulation (\ref{sim}), each step is chosen from a Gaussian distribution centered at the deterministic step in the direction of the force.  
For \emph{crossing events} which climb over a barrier, a trajectory necessarily takes steps in a direction \emph{opposing} the force.
As the center of the Gaussian distribution is always downhill from the present step, the typical step in the \emph{ascent} of a crossing event is in the ``tail'' of the Gaussian distribution (more or less so depending on the parameters entering into the width, $\sigma$).
As the full probability for a crossing event is then a product of steps in the tail of the distribution, it seems clear that only small fluctuations about the most probable crossing event will occur.
Large fluctuations away from the most probable path will be exponentially damped out, as they require steps yet further out in the tail of the distribution.

Our interest in the most probable path for a crossing event, then, is hardly surprising:
to generate an ``important'' (albeit biased) ensemble of trajectories, one must hew to the most probable path.
This we shall do explicitly below, using the results quoted in Sec.\ \ref{sec:mpp}.

But once one knows the most probable path for a single crossing event, the DIMS method still requires the generation of such events at appropriate intervals.
That is, there is an ideal distribution of waiting times between events in a biased simulation (cf. Fig.\ \ref{fig:time}a for the unbiased case).
To see the reason why a distribution of wait times (first passage times) is required, consider Fig.\ \ref{fig:evo}, which demonstrates the rate calculation for biased \emph{and} unbiased simulation.
One first calculates the probability to arrive in the final state (the right well in Fig.\ \ref{fig:potl}), having started in the initial state (the left well), as a function of time.
One then determines the slope of the linear regime.
While an unbiased simulation of sufficient length will automatically generate data with roughly the same precision for all times shown along the horizontal axis in Fig.\ \ref{fig:evo}, a biased DIMS simulation must be designed to do so.
As discussed in \cite{Zuckerman-1999}, one essentially has to evaluate a separate integral for each time point, so part of the goal is to spread the information gathered evenly over the necessary times.
This is the motivation behind the ``thresholding'' detailed below.

\subsection{The Algorithm: Most Probable Crossings above a Threshold}
Specifically, trajectories are generated according to the following algorithm.
Each trajectory is started from the minimum of the left well, $x=-1$, of the potential (\ref{potl}) shown in Fig.\ \ref{fig:potl}, and run for a fixed total amount of time exceeding the transient time, $t_b$.
Pre-defining a threshold value, $-1 < \xth < 0$, of the coordinate $x$,
we perform unbiased simulation while $x < \xth$ according to (\ref{sim}).
If and when the threshold value is exceeded ($x > \xth$), we select steps according to
\begin{equation}
\label{mpcsim}
x_{i+1} \: = \: x_{i} \: + \: \xdot_c(x_{i}) \: \Delta t \: + \: \Delta x_R,
\end{equation}
with $\xdot_c$ given by either the Jacobian-augmented result (\ref{xcomj}) or that without (\ref{xcom}), and with $\Delta x_R$ chosen from the same Gaussian distribution as in the unbiased case.
In other words, instead of using the deterministic step $f_{i} \Delta t/m \gamma$ as in (\ref{sim}), we use the most probable step \emph{in a crossing event} to emulate unbiased crossing events.
The use of a threshold away from the well minimum ensures a sufficiently broad distribution of waiting times between the artificial crossing events, which in turn permits the acquisition of data for the range of times necessary to compute the rate as in Fig.\ \ref{fig:evo}.
We note, however, that it is no more difficult to run a single long trajectory and compute correlation functions to determine rates (e.g., \cite{Pastor-1994b}).

If a Heun scheme \cite{Rumelin-1982,Batrouni-1985,Branka-1998} were used for the unbiased dynamics, the biased dynamics just given (\ref{mpcsim}) could be readily modified in the same way the Euler scheme itself (\ref{sim}) is modified in the Heun approach. 
Given that the underlying continuum action should be the same for the two integration methods, a Heun modification of (\ref{mpcsim}) --- or even the unmodified version --- should give similar results to those found here with the Euler method.

\subsection{Curvature-Adjusted Sampling Width}
Yet another refinement for the one-dimensional biasing techniques is possible, motivated by the time-dependent width of the Gaussian noise, $\sigma_j$, required for the provably optimal computation of the probability \emph{density} at a single point on a curvature-free potential surface \cite{Kloeden-1992}.
Note that the biasing methods just described always used a constant width, 
$\sigma_j = \sigma$ given in (\ref{natural_variance}).
In a spirit similar to Wagner's approach described in \cite{Kloeden-1992}, one can ask the question:
``What is the optimal sampling scheme to travel between two fixed points, with one intermediate step, on a surface with arbitrary curvature?''
Answering this question is not difficult and suggests that the local curvature influences the optimal width.

Following Wagner (see \cite{Kloeden-1992}), one needs to observe first that the optimal \emph{sampling density} at time $t = \dt$ is exactly the distribution of \emph{unbiased} two-step paths which begin at $x_0$ at $t=0$ and end at $x_2$ at $t = 2 \dt$.
This constitutes the perfect \emph{sampling} distribution because it is precisely that (tiny) subset of unbiased trajectories which end at the pre-defined value of interest, and which are distributed naturally.
Using the single-step Gaussian distributions (\ref{trans_prob_density}), the two-step distribution is 
\begin{equation}
\label{two-step}
\bigt(x_2|x_1) \bigt(x_1|x_0) = 
  \frac{1}{2 \pi \sigma} 
  \exp{\left\{ - \frac{  [ (x_2 - x_1) - (f_1/m\gamma) ]^2 
                        + [ (x_1 - x_0) - (f_0/m\gamma) ]^2 
                     }{ 2 \sigma^2} \right\}} \; ,
\end{equation}
where $\sigma$ still represents the unbiased value and $f_i$ again gives the force at $x_i$.
By rearranging the terms, completing the square and approximating
$f_i \simeq f_0 - (x_i - x_0) \, U''(x_0)$, one finds
\begin{equation}
\label{curv-dist}
\bigt(x_2|x_1) \bigt(x_1|x_0) \simeq c \times
  \exp{\left\{ - \frac{ \left[ x_1 - \frac{x_0+x_2}{2}\left( \frac{1-\alpha}{1-\alpha+\alpha^2/2} \right)\right]^2
                     }{ 2 \left[\sigma/\sqrt{2(1-\alpha+\alpha^2/2)}\right]^2} \right\}}
\; ,
\end{equation}
where $c$ is a constant independent of $x_1$ in this approximation, and the dimensionless curvature is $\alpha \equiv U''(x_0) \dt/ m \gamma$.

The result (\ref{curv-dist}) for the optimal distribution of $x_1$ values illustrates several interesting points.
First, the distribution is independent of the force to first order in $\alpha$.
Second, while the correction to the expected mean of the distribution, $(x_0 + x_2)/2$, is second order in $\alpha$, that for the width, $\sigma$, is first order --- noting that the factor $\sqrt{2}$ is expected from the work of Wagner and actually approaches unity for a large number of time steps \cite{Kloeden-1992}.
The optimal sampling width in fact \emph{decreases} at the barrier top, where $\alpha < 0$, in the small $\dt$ limit ($\alpha \gg \alpha^2$).
Thus, the distribution (\ref{curv-dist}) motivates a further bias for sampling, namely use of the width
\begin{equation}
\label{sigma-samp}
\sigma_{\mathrm{samp}} = \sigma/\sqrt{1-\alpha+\alpha^2/2}
\end{equation}
for sampling from Gaussian distributions near the point $x$, where $\alpha$ is to be computed using $U''(x)$.
We investigate this refinement in the next section.

\pagebreak
\begin{figure}
\begin{center}
\epsfig{file=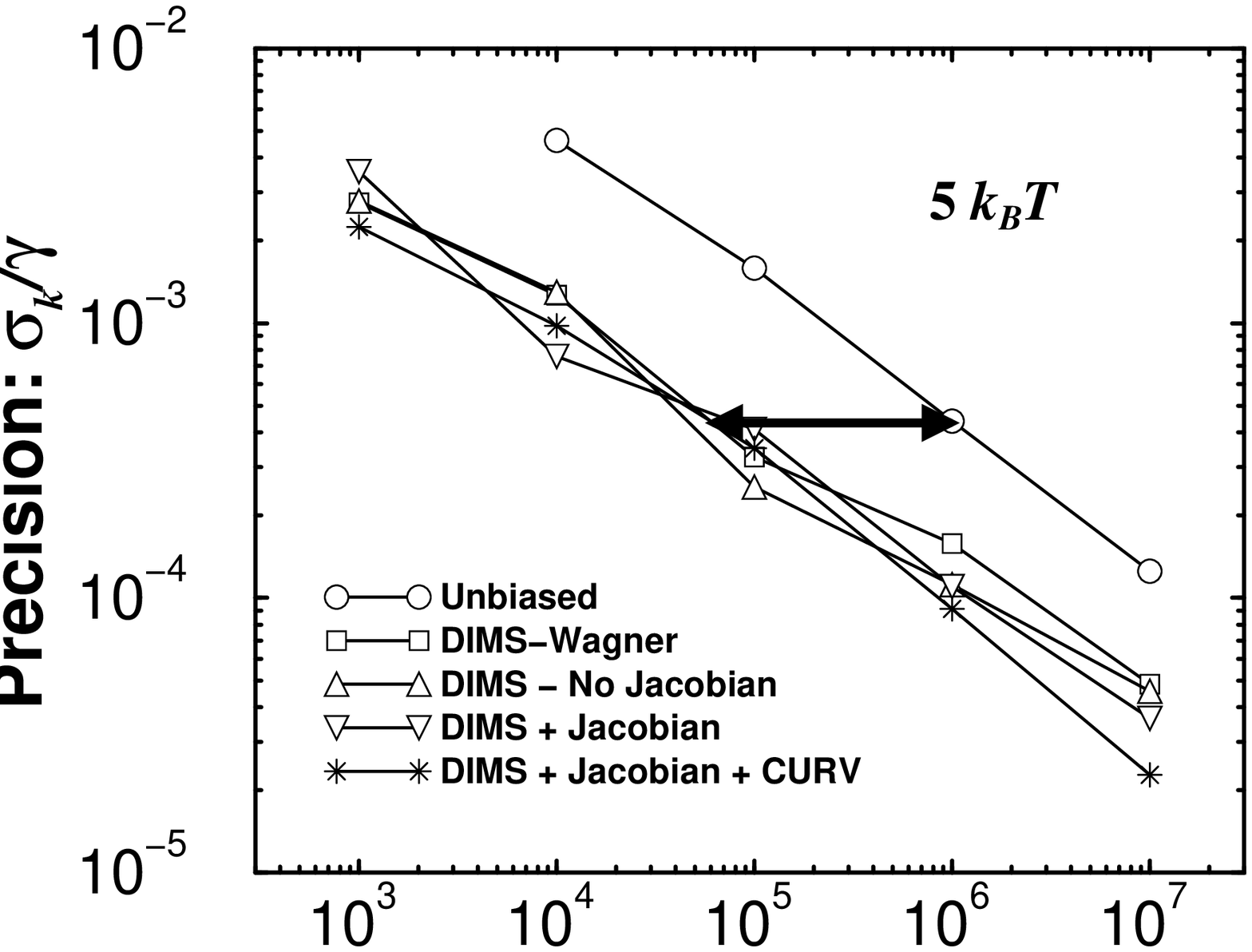, height= 2.75in}

\epsfig{file=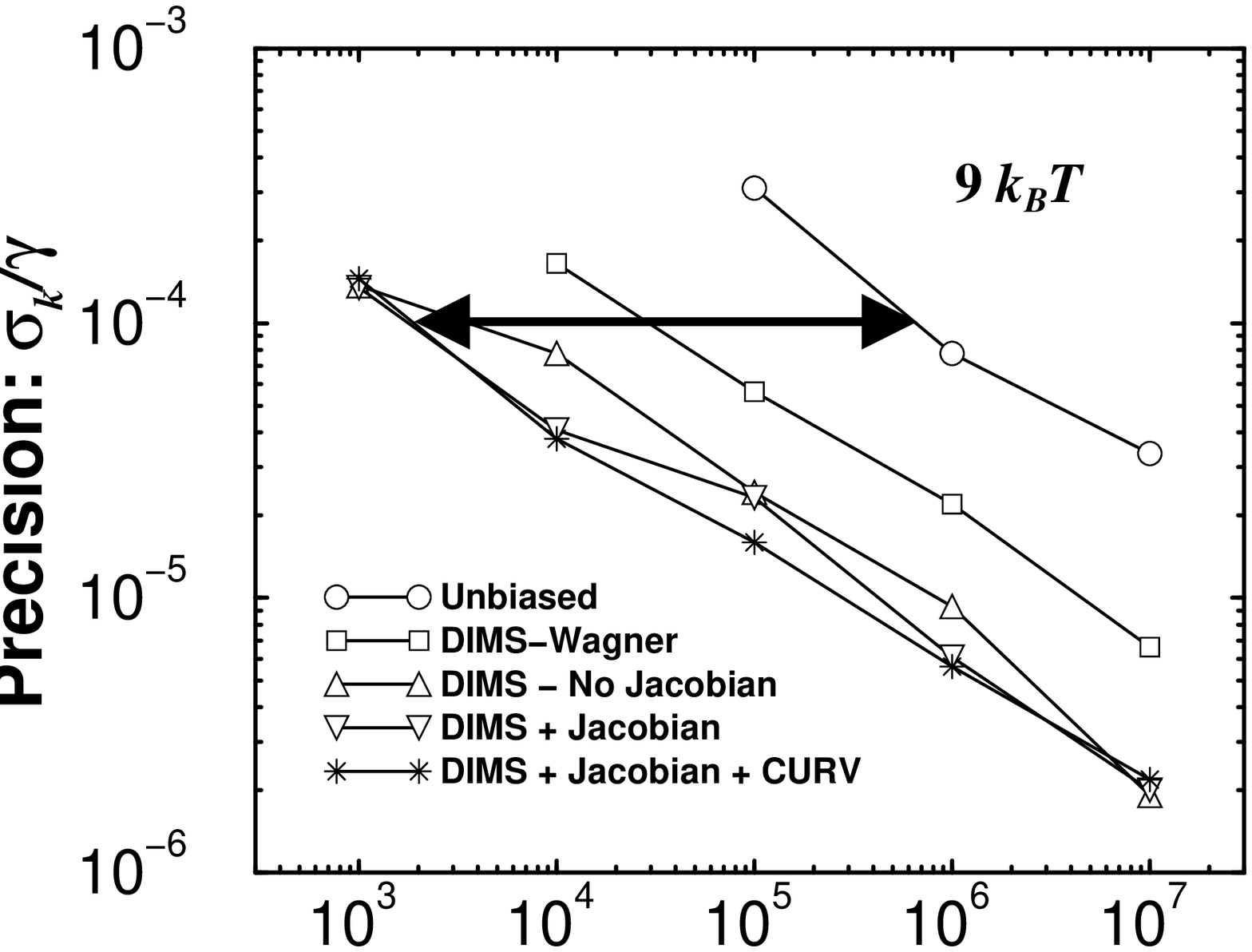, height= 2.75in}

\hspace*{-0.53cm}\epsfig{file=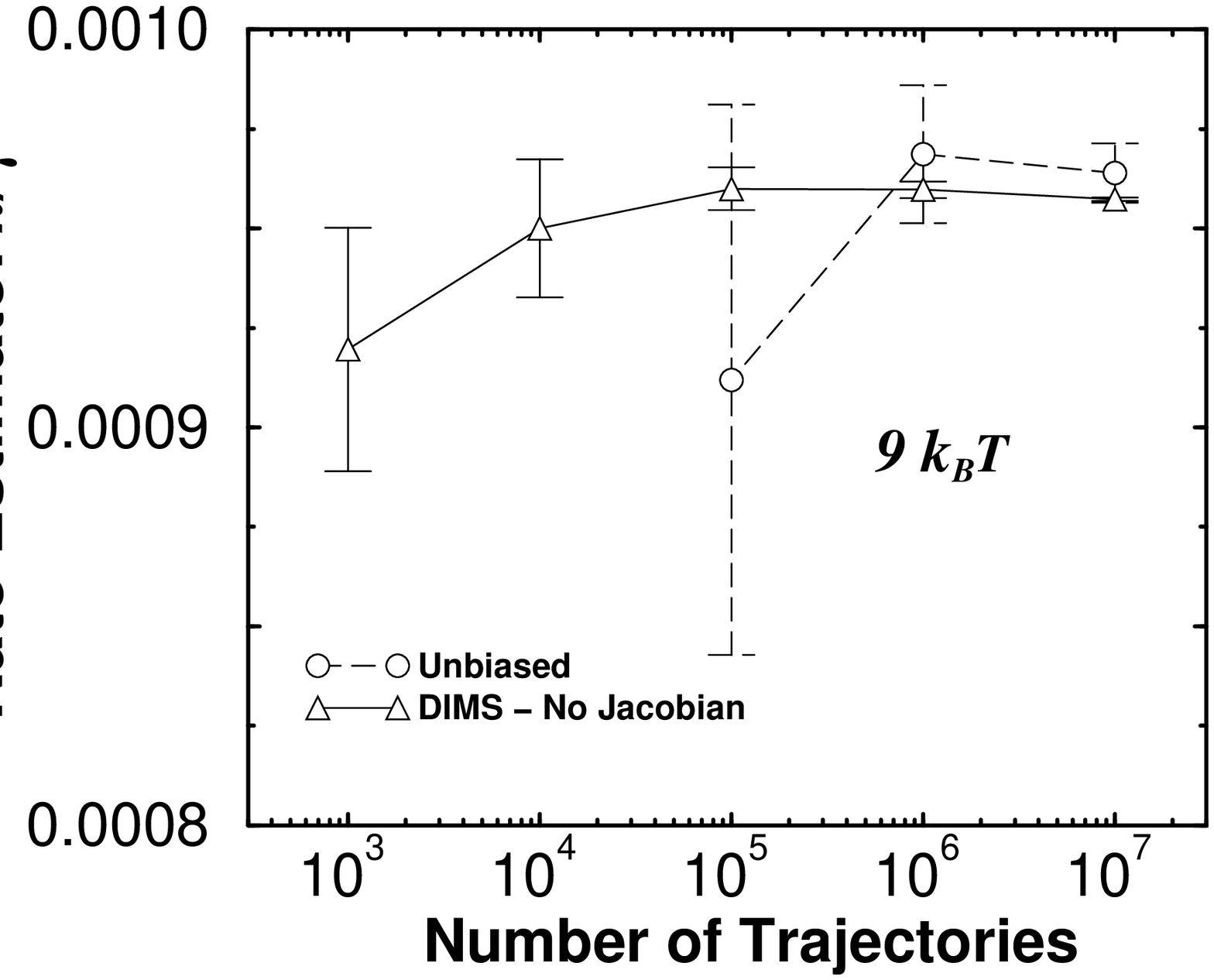, height=3.04in}
\end{center}
\caption{\label{fig:effic}
Efficiency in Rate Computations using the DIMS approach.
The standard deviation, $\sigsim$, of a set of 20 rate estimates computed by either unbiased or DIMS simulation is plotted against the length of each simulation (top 2 plots).
The spans of the horizontal arrows measure the efficiency by indicating the differences in simulation length required to obtain a given variance (i.e., precision).
For example, when the barrier height is $E_b = 9 k_B T$, the best DIMS approaches are roughly 300 times faster in obtaining a precision given by $\sigsim = 10^{-4}$.
The ``DIMS-Wagner'' algorithm is from our earlier work [14],
while the other DIMS simulations use the algorithm of Eq.\ (\ref{mpcsim}) and the surrounding text.
For the latter, the most probable step is chosen according to either the Jacobian-augmented formulation (\ref{xcomj}) or that without (\ref{xcom}), as indicated.
The label ``CURV'' indicates that the Gaussian sampling width was modified according to (\ref{sigma-samp}).
The bottom plot shows rate estimates, $k$, for the $9 k_B T$ barrier height.
Both DIMS [Eqs.\ (\ref{mpcsim}) and (\ref{xcomj})] and unbiased estimates converge toward a common result with increasing simulation length.
The error bars indicate the standard error of the mean, and under-estimate the 95\% confidence interval.
}
\end{figure}
\pagebreak
\section{Results: Comparison of Efficiency}
\label{sec:results}
We demonstrate the capability of the DIMS algorithms of the previous section by quantifying their efficiency for rate computations in the simple bistable potential shown in Fig.\ \ref{fig:potl},
\begin{equation}
\label{potl}
U(x) = E_b \left( (x/l)^2 - 1 \right)^2,
\end{equation}
where $E_b$ is the barrier height and $l$ is the lengthscale of the problem.
The central result is that one must account for the most probable crossing to gain maximum speed-up in rate computation as compared with unbiased simulation.
Not surprisingly, the efficiency increases with barrier height.
Yet even for the relatively low barrier height of $E_b = 5 k_B T$, we achieve roughly a 20-times efficiency improvement --- i.e., DIMS rate calculations are approximately 20 times as fast for a given level of precision.
That factor increases to 300 for a $9 k_B T$ barrier.
While such gains will not be readily extendible to multi-dimensional systems, it is important to understand and demonstrate the ingredients necessary for optimal performance.

Fig.\ \ref{fig:effic} shows our results for the potential (\ref{potl}) for two different barrier heights, $E_b/k_B T = 5$ and 9.
The biasing methods accounting for the most probable crossing are significantly superior for the larger barrier.
The ``DIMS-Wagner'' algorithm --- which takes no account of the most probable path --- refers to the technique described in Ref.\ \cite{Zuckerman-1999}.
It performs only modestly well for the $9 k_B T$ barrier, and its efficiency is very sensitive to the fixed simulation length.
All the other DIMS procedures employed the algorithm of the previous section, (\ref{mpcsim}), above a threshold $\xth = -0.7$, with trajectories initiated at $x=-1$.
The presence or absence of the ``Jacobian'' (stochastic correction term) reflects whether (\ref{xcomj}) or (\ref{xcom}) was used to complete (\ref{mpcsim}),
and ``CURV'' indicates that the curvature-modified width (\ref{sigma-samp}) was used in place of the unbiased width (\ref{natural_variance}).
Rates, $k$, were calculated using the method noted in Sec.\ \ref{sec:mot} and Fig.\ \ref{fig:evo}.
Efficiency is computed by estimating the relative simulation lengths needed to obtain a desired degree of precision, given by the variance of the rate estimates,
\begin{equation}
\label{sigrate}
\sigsim^2 = \frac{1}{n} \sum_{i=1}^n \left( k_i - \langle k \rangle \right)^2,
\end{equation}
where $n=20$ is the number of simulations performed for each data point, $k_i$ is the rate computed for the $i$th simulation, and $\langle k \rangle$ is the mean of the $n$ rate estimates.
The horizontal arrow spanning the DIMS and unbiased results for $E_b = 9 k_B T$
at a precision of $\sigsim = 10^{-4}$, for example, exceeds two decades --- indicating that the DIMS computations are more than 100 times faster.

The effectiveness of the DIMS formulation \emph{excluding} the Jacobian --- i.e., based on (\ref{mpcsim}) and (\ref{xcom}) with $C=0$ --- deserves further comment.
While the Jacobian-augmented DIMS simulation is slightly superior for $E_b = 9 k_B T$, the insensitivity to including the correction term is surprising given the sharp contrast demonstrated in Fig.\ \ref{fig:cross_eb7}.
The lesson appears to be that, at least for the parameters studied, the motion in the immediate neighborhood of the barrier top (where the correction term, proportional to the curvature, has greatest effect) is less important to the anatomy of a crossing event than the (rapid) climb and descent.
In the long term, the success of the uncorrected approach could facilitate the extension of DIMS to multi-dimensional systems, since that approach does not require computation of second derivatives of the potential.
Future work may show this to save a substantial amount of computer time.

We note that our efficiency estimates have excluded the ``overhead'' cost of implementing the DIMS method.
This cost depends on the optimization of one's code, and as it happens, our code is sub-optimal for \emph{un}biased simulation, so that there is no overhead at all.
There are, however, inherent overhead costs in DIMS that cannot be optimized away.
While the dynamics employing the Jacobian-augmented most probable path (\ref{xcomj}) requires the computation of a second derivative at every step, our results show that the simpler form (\ref{xcom}) is nearly as good and requires only the force.
Calculating the force, it should be remembered, is not an overhead cost because this must be done in unbiased simulation (\ref{sim}) anyway.
The only notable cost inherent in the DIMS method, then, is computing the error associated with biased computation --- in order to correct for it as discussed in Refs.\ \cite{Woolf-1998e,Zuckerman-1999}.
This correction entails computing the ratio of two Gaussian terms (or, equivalently, the difference of two logarithms) at every step.
Compared with the fixed cost of unbiased simulation --- computing the force and generating a high quality pseudo-random number at every step --- and its inherent inefficiency with long waiting times, the DIMS costs are far overshadowed by the efficiency gains which here exceed one and two orders of magnitude.

For completeness, we give a number of further details, which apply to both the unbiased and DIMS results.
A simulation consisted of $N$ trajectories (see Fig.\ \ref{fig:effic}) initiated at $x=-1$ at time $t=0$.
The time steps were $\Delta t = 0.003 \gamma^{-1}$ for $E_b = 5 k_B T$ (having set $\gamma \equiv 1$) and 
$\Delta t = 0.001 \gamma^{-1}$ for $E_b = 9 k_B T$.
These were determined to be close to the maximum values for which the rate estimates did not change as $\dt$ increased in unbiased simulation.
As discussed above, the rate is computed as the slope of the linear regime in a plot of the arrival probability (to be in the right well, $x>0$) as a function of time;
see Fig.\ \ref{fig:evo}.
The slopes (rates, $k_i$) were computed from a least-squares fit to 10 data points, the $t$ values of which were held fixed for a given barrier height.
The particle mass, $m$, and the thermal energy, $k_B T$, were both set to 1.
Note that Fig.\ \ref{fig:cross_eb7} uses the parameters given in the caption.

\subsection{What is the Optimal Efficiency?}
``The system is so simple.  How does one know whether a 300-time improvement in efficiency is impressive?''
So a skeptic might wonder, and the attempt to answer seems a worthwhile exercise.

The basic point is that computing a rate by simulation involves the simultaneous calculation of a series of difficult integrals of the form (\ref{pbint}) discussed in Sec.\ \ref{sec:back}.
Each data point in Fig. \ref{fig:evo} is an estimate for such an integral.

We can try to estimate the minimum number of trajectories needed for the rate computation by multiplying together estimates for the following:
(i) the number of trajectories needed to estimate the probability \emph{density} to be at a single location, $x$, in state B at a given time, $P(t_j; x \in B | 0; A)$;
(ii) the number of discrete locations $x$ in state B required to estimate the probability for the whole state; and,
(iii) the number of independent time points needed to estimate a slope in the linear regime.
Regarding (i), only in the case of a constant force can $P(t_j; x \in B | 0; A)$ be computed exactly \cite{Kloeden-1992} --- equivalently, with a single trajectory.
One might expect that at least 10 trajectories would be required for any real surface, setting (i).
In a similar manner for (ii), at least 10 points should be required to characterize a state (which, in principle, is known only numerically).
The number of independent time points is a slightly more complex issue since some (though not all) trajectories from a given time point may also be used to estimate another.
Conservatively, then, we use the estimate three for (iii).
Our estimate for the minimum number of trajectories required to calculate a rate is thus 300.
We believe our DIMS results of Fig.\ \ref{fig:effic} compare favorably with this heuristic --- and conservative --- theoretical minimum.

\pagebreak
\section{Possible Extensions to Multi-dimensional Problems}
\label{sec:multi}
While the results presented here for a one-dimensional potential seem a far cry from a high-dimensional biomolecular system, we believe they teach important lessons for the large system.
The simplest point is that a biased trajectory must closely mimic the natural barrier-crossing dynamics if efficiency in rate computations is to be obtained.
Indeed, we have found that a poor bias can be worse than no bias at all.
It is not enough to know --- as one does automatically in one dimension --- even the optimal geometric channel for a transition:
the size of the steps along that geometric path are critical, as we have shown.

Although the problem of finding channels is extremely challenging in itself, let us ask ``How can one compute the rate for a large system assuming the channels are known?''
A natural choice would be to start with an initial trajectory within each channel, and then attempt either to optimize it \cite{Berkowitz-1983,Elber-1990,Elber-1991,Karplus-1992,Theodorou-1993,Elber-1997,Elber-1996,Elber-1999a} or to generate an ensemble of trajectories from it \cite{Chandler-1998a,Eastman-2000}.
Given the importance of closely following the optimal course,
it seems natural to use an optimization or sampling scheme which builds in the knowledge of the most probable path (\ref{xcom}) and its multi-dimensional analog.
The risks, otherwise, could be great:
since a high-dimensional path will be very rough, it is easy to imagine a multi-step segment of a trajectory becoming trapped in a region of the potential surface with far too few or too many steps to be even close to optimal.
One idea for overcoming this difficulty would be to use a scheme capable of removing and inserting time steps, in order to search for an appropriate distribution of steps along a pre-defined geometric path.
We intend to pursue further investigations along these lines.

Returning to the issue of finding multi-dimensional channels in the first place, we note that the DIMS method is ideally suited to attack this problem since it generates an ensemble of completely \emph{independent} trajectories.
Indeed, we have already developed an algorithm which has proved capable of efficiently finding distinct, important channels in a multi-dimensional system \cite{Zuckerman-2000b}.
We have named the idea the ``soft-ratcheting algorithm,'' and we note that its efficiency is thus far limited to finding channels, rather than determining rates.
The essence of the technique is simple: generate an unbiased step and accept it with a \emph{probability} (hence the ``softness'') depending on how far the trajectory has progressed toward the target state.
To complete the calculation in the DIMS formulation, one then estimates the overall acceptance probability --- which is an inexpensive calculation in a large system.
Note that the soft-ratcheting algorithm requires no second derivatives of the potential.

Finally, a timescale problem could prove serious, even though we do not expect it to be nearly the handicap it is for molecular dynamics.
In particular, a fundamental limitation of applying the DIMS method (or a related approach \cite{\dimscite}) to multi-dimensional problems is the barrier-crossing time, $t_b$.
In practical terms, $t_b$ shows up as the transient time prior to the linear regime in a plot used for rate evaluation (Fig.\ \ref{fig:evo}).
Probability cannot accrue, after all, until crossing events have occurred.
In a large system, $t_b$ is the limiting timescale for applying a method like DIMS to computing rates.
Since a reasonable number, say $N$, crossing events will be needed to estimate the rate, one would have to simulate for a time exceeding $N t_b$.
The authors are unaware of any estimates of $t_b$ for biomolecular systems, but we note that --- for the DIMS method to potentially yield a rate estimate --- $t_b$ would have to be less than a nanosecond for a large, explicitly-solvated protein.
Recent work on large time steps \cite{Gillilan-1992,Elber-1996,Elber-1999a,Eastman-2000} holds promise, however, for attacking the timescale problem.


\pagebreak
\section{Summary and Conclusions}
\label{sec:sum}
We have demonstrated substantial increases in efficiency for simulation-based calculations of transition rates in bistable potentials with modest barrier heights, $5 k_B T$ and $9 k_B T$, using new biasing methods within the dynamic importance sampling (DIMS) formulation \cite{Woolf-1998e,Zuckerman-1999}.
Computations were sped up by a factor of approximately 300 for the $9 k_B T$ barrier, and the primary results (Fig.\ \ref{fig:effic}) suggest the speed-up will increase significantly for larger barriers.
The critical ingredient in our efficiency was close emulation of probable crossing events suggested by the Onsager-Machlup action, (\ref{action}) and (\ref{som}).

The simple one-dimensional problem has been addressed from a variety of theoretical and numerical perspectives in an effort to pave the way for more difficult problems.
In Sec.\ \ref{sec:mpp} we examined the stochastic correction --- or ``Jacobian'' --- term (\ref{somj}) in the Onsager-Machlup action from theoretical and numerical perspectives.
There, we also discussed the desired distribution of waiting times between artificial crossing events, as well as the effect (and utility) of the curvature of the potential.
After presenting the explicit results for efficiency levels in rate computations, we discussed the optimal efficiency one could hope to attain in principle.

In commenting on the extension of the DIMS method to large, high-dimensional systems in Sec.\ \ref{sec:multi}, we noted that the problem may be conceptually broken up into two parts:
finding the geometric channels, and then sampling trajectories within those channels.
The DIMS method is ideally suited for the first step, channel-finding, and we described an explicit algorithm effective in that capacity.
Our hope is that the results of the present paper will be useful in constructing techniques for the second stage, single-channel trajectory sampling.


\pagebreak
\begin{acknowledgments}
Many people generously offered comments on early versions of the manuscript:
Ron Elber, Alan Grossfield, Christopher Jarzynksi, Rohit Pappu, Horia Petrache, Jonathan Sachs, David Shalloway, and Katharina Vollmayr-Lee.
The authors also thank David Chandler for informative discussions, Phillip Geissler for pointing out Ref.\ \cite{Mortensen-1969}, and
the referee for a number of informative comments, particularly regarding Sec.\ \ref{sec:mpp}.
Shlomo Raz kindly provided additional computing facilities for this work.
We gratefully acknowledge funding provided by the NIH (under grant GM54782), the AHA (grant-in-aid), the Bard Foundation, and the Department of Physiology.
\end{acknowledgments}

\bibliographystyle{prsty}

\end{document}